\documentclass[namedreferences]{solarphysics}
\usepackage[natbib,optionalrh,solaromanenum]{spr-sola-addons} 
\usepackage{graphicx}                    
\usepackage{color}                       
\usepackage{url}                         

\usepackage{hyperref}

\renewcommand{\vec}[1]{{\mathbfit #1}}

\begin{document}

\begin{opening}

\title{On the Pulse Shape of Ground Level Enhancements}

%
\author{R.D.~\surname{Strauss}$^{1}$\sep
		O.~\surname{Ogunjobi}$^{1}$\sep 
        H.~\surname{Moraal}$^{1}$\sep
        K.G.~\surname{McCracken}$^{2}$ \sep      
		R.A.~\surname{Caballero-Lopez}$^{3}$
       }
\runningauthor{R.D. Strauss {\it et al.}}
\runningtitle{On the Pulse Shape of Ground Level Enhancements}
%
  \institute{$^{1}$ Centre for Space Research, North-West University, Potchefstroom, South Africa.\\
                     Email: \url{dutoit.strauss@nwu.ac.za}\\
$^{2}$ Institute for Physical Science and Technology, University of Maryland, College Park, MD, 20742, U.S.A.\\
$^{3}$ Ciencias Espaciales, Instituto de Geof\'isica, Universidad Nacional Aut\'onoma de M\'exico, 04510 M\'exico D.F., M\'exico. \\
             }

\begin{abstract}

We study the temporal intensity profile, or pulse shape, of cosmic ray ground level enhancements (GLEs) by calculating the rise $(\tau_\mathrm{r})$ and decay $(\tau_\mathrm{d})$ times for a {small} subset of all available events. Although these quantities show very large inter-event variability, a linear dependence of $\tau_\mathrm{d} \approx 3.5 \tau_\mathrm{r}$ is found. We interpret these observational findings in terms of an interplanetary transport model, thereby including the effects of scattering (in pitch-angle) as these particles propagate from (near) the Sun to Earth. It is shown that such a model can account for the observed trends in the pulse shape, illustrating that interplanetary transport must be taken into account when studying GLE events, especially their temporal profiles. Furthermore, depending on the model parameters, the pulse shape of GLEs may be determined entirely by interplanetary scattering, obscuring all information regarding the initial acceleration process, and hence making a classification between impulsive and gradual events, as is traditionally done, superfluous. 

\end{abstract}

%
\keywords{Cosmic Rays: Solar; Energetic Particles: Propagation; Magnetic fields: Interplanetary}

\end{opening}

%

\section{Introduction}

Ground level enhancements (GLEs) are sudden increases in the cosmic ray (CR) intensity as observed at Earth's surface,  in recent times mostly through neutron monitors (NMs). Since the first observation of a GLE on 28 February 1942, such events have been observed seventy-one times \citep[see the review by][]{McCraken:2012}. {A database of all GLE events is available at \url{ftp://cr0.izmiran.rssi.ru/COSRAY!/FTP_GLE/} (hosted by E. Eroshenko), \url{https://gle.oulu.fi/}, as well as \url{http://usuarios.geofisica.unam.mx/GLE_Data_Base/} with details of the latter described by \citet{MoraalandRogelio:2014}.} It is generally believed that transient solar eruptive episodes, such as coronal mass ejections (CMEs) and solar flares, are responsible for producing the enhanced CR flux detected during a GLE event, and hence these particles can be classified as solar energetic particles (SEPs). If the primary SEPs reach Earth with kinetic energies $\geq $ 500 MeV per unit charge, or rigidity $\geq$ 1 GV, they can produce atmospheric showers of secondary particles that can reach ground level and be registered by NMs. {See also the work of, amongst others, \citet{Shea:1994}, \citet{Miroshnichenkoetal2000}, \citet{Ohetal:2012}, and \citet{rogeliomoraal2016}.} \\

\begin{figure*} 
\centerline{\includegraphics[width=0.99\textwidth,angle=0,clip=]{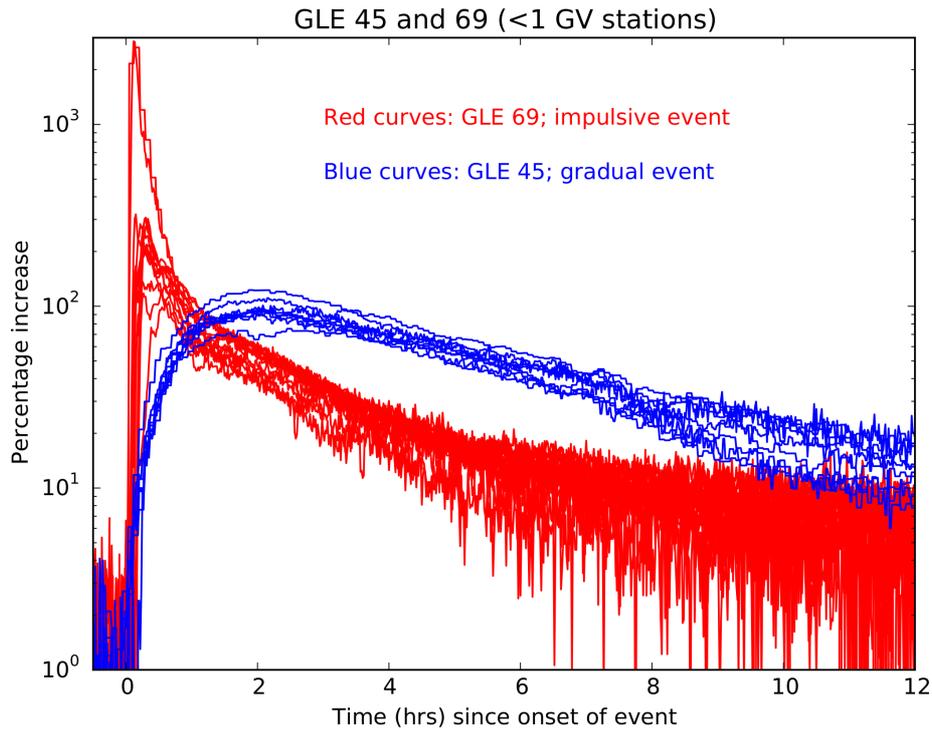}}
\caption{The temporal profile, or pulse shape, of GLE 45 (blue curves) and GLE 69 (red curves) in order to illustrate the classification of so-called gradual (such as GLE 45) and more impulsive (such as GLE 69) events. {For each event, the different lines correspond to measurements by different NMs.}}\label{fig_compare_GLEs}
\end{figure*}

Despite the large amount of SEP observations by both ground and space based CR detectors over the years, the relationship between flares and CMEs, and their role in accelerating particles to relativistic energies during major solar events remains an exigent scientific challenge. Most studies, {\it e.g.} \citet{Reames:1999}, classify SEPs into two distinct classes, impulsive and gradual events. Particles from these events are believed to be accelerated by different processes in different transient structures; impulsive SEPs are associated with magnetic reconnection in solar flares, while gradual SEPs are thought to be accelerated mainly by diffusive shock acceleration in the shock fronts of large CMEs. Having an accurate classification of SEP events may therefore give insight into where and how these particles were accelerated \citep[\emph{e.g.}][]{Reames:2013}. The usual classification, based on the shape of the observed intensity time profile, the so-called pulse shape, is furthermore motivated by differences in the composition and charge state of the SEPs \citep[\emph{e.g.}][]{Mewaldt:2012}, for example, impulsive events have higher ionization states. For ground-based observations of GLE, however, both the composition and change state information is lost and the GLEs are classified exclusively on their pulse shape. In Figure \ref{fig_compare_GLEs}, as an example, we show the pulse shape of GLE 45 (blue curves) and GLE 69 (red curves) for all available NM stations with a cut-off rigidity below 1 GV. It is clear that GLE 45 was observed to be more gradual than the more impulsive GLE 69, but it is impossible to extract information solely from this temporal profile about the acceleration mechanisms responsible for each particle event without taking interplanetary transport into account. Once the SEPs are accelerated, presumably near the Sun, they still have to propagate (mainly) along the turbulent interplanetary magnetic field to reach Earth. Depending on the level and nature of the underlying turbulent fluctuations and how efficiently they can disrupt ({\it i.e.} scatter) the particle trajectories, the SEP temporal profile reaching Earth may be significantly different from that near the acceleration region.\\

In this study we will show that the interplanetary transport conditions may alter the pulse shape in such a way as to remove any source information by the time it reaches Earth, possibly making the classification of GLEs into different classes superfluous. \\ 

\section{Characterising the Pulse Shape}
\label{Sec:charactering_the}

 \begin{table}
 \begin{minipage}{\textwidth}    
 \caption{A summary of the GLE events selected for this study. The approximate solar longitude of the transient event and the maximum increase is taken from \cite{McCraken:2012}.} \label{Tab:events_table}
 \begin{tabular}{cccccc}     
 \hline

GLE no. & Sol. long.\footnote{Ideal magnetic connection between the source of the SEPs and Earth would correspond to a source located at $\approx 60^{\circ}$ W in terms of solar longitude.} & Max. increase \footnote{The maximum increase of the omni-directional counting rate.}   & Rise time & Decay time & NM stations \footnote{ALTR -- Alert, APTY -- Apatity, BRBY -- Barentzburg, CALG -- Calgary, CAPS -- Cape Schmidt, DPRV -- Deep River, FSMT -- Fort Smith, GSBY -- Goose Bay, INVK -- Inuvik, MCMD -- McMurdo, MRNY -- Mirny, MWSN -- Mawson,  NRLK -- Norilsk, OTWA -- Ottawa, PWNK -- Peawanuk, TERA -- Terra Adelie, THUL -- Thule, TXBY -- Tixie Bay} \\
 & ($^{\circ}$ W) & (\%)   & (min) &  (min) &  ($< 1$ GV)\\
 \hline
 30 & 40 & 19 & 25.2 & 75 & ALRT, APTY, DPRV, GSBY, INVK,  \\
 	& & & & & MCMD, MRNY, OULU, TXBY, SANAE\\ 

 31 & 68 & 28 & 5.4 & 19.8 & ALRT, APTY, DPRV, GSBY, INVK, \\
 	& & & & & MCMD, OULU, TXBY,  SANAE\\ 

 38 & 86 & 17 & 16.2 & 84.6 & ALRT, APTY, DPRV, GSBY, INVK,   \\
 	& & & & & MSWN, OULU, TERA, THUL, TXBY, \\ 
	& & & & &  SANAE, CALG\\ 

 39 & 130 & 30 & 6 & 16.2 & ALRT, APTY, DPRV, GSBY, INVK,   \\
 	& & & & & MCMD, MWSN, OULU, TERA, THUL,   \\
 	& & & & & TXBY, SANAE, CALG\\ 

 42 & 105 & 246 & 78 & 158.4 & APTY, CALG, CAPS, DPRV, INVK,  \\
 	& & & & & MCMD, MWSN, MRNY, OTWA, OULU, \\
 	& & & & & TERA, THUL, TXBY, SANAE\\ 

 43 & -9 & 41 & 69 & 443.4 & CALG, CAPS, DPRV, GSBY, INVK, \\
 	& & & & & MCMD, MWSN, OTWA, OULU, TERA, \\
 	& & & & & THUL,  SANAE \\ 

 45 & 57 & 91 & 75.6 & 241.8 & APTY, CALG, CAPS, DPRV, GSBY,  \\
 	& & & & & INVK, MCMD, MWSN, OTWA, OULU, \\
 	& & & & & TERA, THUL, SANAE \\ 

 48 & 76 & 11 & 6 & 75 & CALG, CAPS, DPRV, GSBY, INVK, \\
 	& & & & & MWSN, MCMD, OTWA, OULU, THUL, \\
 	& & & & & TXBY, SANAE \\ 

 52 & 70 & 21 & 25.2 & 80.4 & APTY, CALG, DPRV, GSBY, INVK, \\
 	& & & & & MCMD, OULU, TERA, THUL, TXBY, \\
 	& & & & & SANAE \\ 

 59 & 7 & 31 & 22.8 & 96 & APTY, CALG, GSBY, INVK, MCMD, \\
 	& & & & & MSWN, OULU, THUL, TXBY, SANAE \\

 60 & 85 & 80 & 25.8 & 73.2 & APTY, CALG, CAPS, FSMT, MCMD, \\
 	& & & & &  NAIN, NRLK, OULU, PWNK, TERA, \\
 	& & & & &  THUL, TXBY,  SANAE\\ 

 65 & -2 & 12 & 22.8 & 40.2 & APTY, BRBY, CAPS, FSMT, INVK,  \\
 	& & & & &  MCMD, NAIN, NRLK, OULU, PWNK, \\
 	& & & & &  TERA, THUL, TXBY, SANAE\\ 

 69 & 58 & 430 & 3 & 16.2 & APTY, BRBY, CALG, CAPS, FSMT, \\
 	& & & & &  INVK, MCMD, MWSN, NAIN, NRLK, \\
 	& & & & &  OULU, TERA, THUL, TXBY,  SANAE\\ 

 70 & 24 & 27 & 33.6 & 70.2 & APTY, BRBY, CAPS, FSMT, INVK,  \\
 	& & & & &  MCMD, MWSN,  NAIN, NRLK, OULU,    \\
 	& & & & &  PWNK, TERA, THUL, TXBY,  SANAE\\ 

 \hline
 \end{tabular}
 \end{minipage}
 \end{table}

\begin{figure*} 
\centerline{\includegraphics[width=1\textwidth,angle=0,clip=]{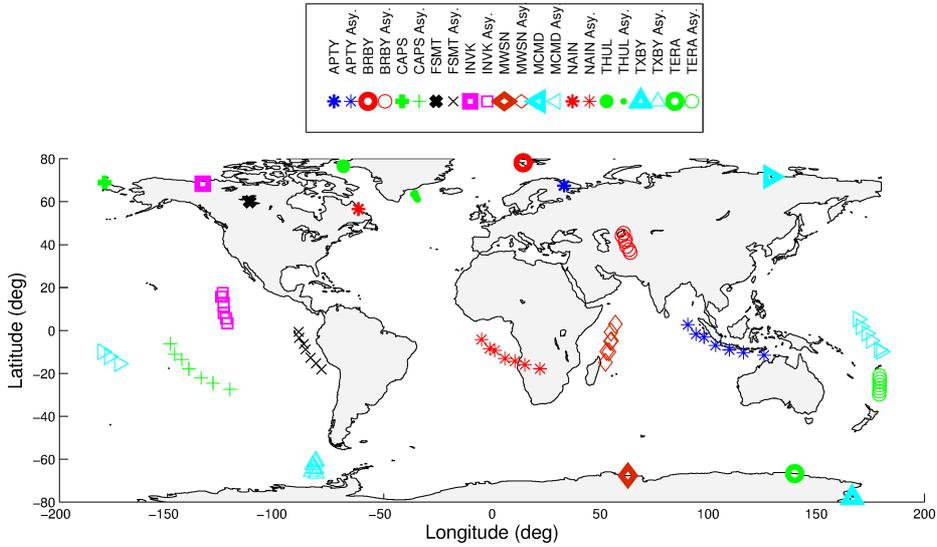}}
\caption{Geographical locations of a selection of the contributing NMs with geomagnetic cut-off rigidity below 1 GV (large markers) and the corresponding asymptotic viewing directions (small markers). For each NM, the asymptotic viewing direction is calculated for rigidities of 0.7, 1.4, 2.1, 2.9,  3.6, 4.3 and 5.0 GV, with the highest rigidity position closest to each station.}\label{fig_viewing_angle}
\end{figure*}

The data base, which contains all available NMs observation, of all 71 GLEs since 1942, has been decsribed by \cite{McCraken:2012}. In this study, we select only 14 GLEs, based on the following 5 criteria: i) in order to remove any energy effects, we only use NMs with cut-off rigidities below 1 GV{, as all these NMs are sensitive to the same particle energy range. As we will show later in this paper, particle propagation is very dependent on the transport coefficients, which are believed to be energy dependent, making a direct comparison of the pulse shape of different energy particles (i.e. NMs with different cut-off rigidities) impossible.} ii) We select only GLEs that were observed simultaneously by more that 10 NM stations. It is well-known that NMs are sensitive to the arrival direction of particles due to deflection by the Earth's magnetosphere \citep[\emph{e.g.}][]{Shea:1990}. For an anisotropic event, such as a GLE where particles are predominantly streaming towards Earth, different stations at the same cut-off rigidity may respond differently to the same SEP pulse. Note, for example, the difference in peak intensity of GLE 69 (shown in Figure \ref{fig_compare_GLEs}) as observed by different NMs due to this geographical effect of essentially viewing particles with different pitch-angles. By using more NM stations, we try, as far as possible, to remove this observational bias. In Figure \ref{fig_viewing_angle} we show the calculated asymptotic viewing directions for a number of NM stations used in this study. Note the nearly uniform longitudinal coverage near the equatorial regions. iii) The selected GLEs have amplitudes more than 10\% and iv) consists of both small and large events v) showing both gradual and impulsive characteristics. Table \ref{Tab:events_table} summarizes all the events that were selected and used in this study. We consider this selection as a good subset of all available 71 events. Lastly, before quantifying the pulse shape, we average the time profiles of the separate NMs to remove the observational pitch-angle bias and refer to this as the omni-directional time profile. This is also done in order to facilitate a later comparison to modelled solutions.\\

\begin{figure*} 
\centerline{\includegraphics[width=0.99\textwidth,clip=]{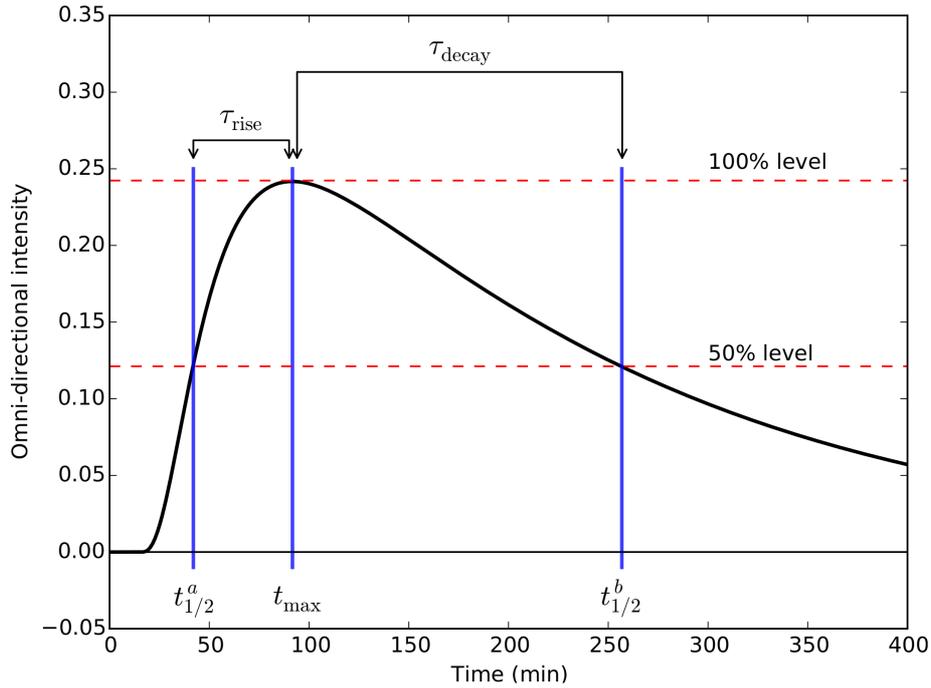}}
\caption{Illustrating how we characterize the pulse shape of a GLE through the rise and decay times.}\label{fig_define_times}
\end{figure*}

In Figure \ref{fig_define_times}, we demonstrate how the pulse shape is characterized in this study: for each event, the time profile of the omni-directional intensity is calculated and the time of maximum intensity, $t_{\max}$, recorded. We then search for times when the intensity is half of its maximum value and note the corresponding times $ t^a_{1/2}$ and $t^b_{1/2}$. The rise time is now defined as $\tau_\mathrm{r} := t_{\max} - t^a_{1/2}$ and the decay time as $\tau_\mathrm{d} := t^b_{1/2} - t_{\max}$. Here we deviate from \citet{moraaletal2016}, and usual SEP studies such as \citet{dresingetal2014}, by not calculating the so-called onset time (defined as the time when the observed intensity breaches the background level) and defining the rise time in terms of this quantity. Observationally, the onset-times are biased towards larger events, or events with fast rising intensities ({\it i.e.} for impulsive events), that breach the background quicker \citep[\emph{e.g.}][]{xieetal2016}, while, from a modelling point-of-view any chosen background level in a model is somewhat arbitrary \citep[\emph{e.g.}][]{wangqin2015}. Our re-definition of the rise time removes this bias, while sufficiently quantifying the temporal slope  of the increasing phase of an SEP event.\\

\begin{figure*} 
\centerline{\includegraphics[width=0.99\textwidth,clip=]{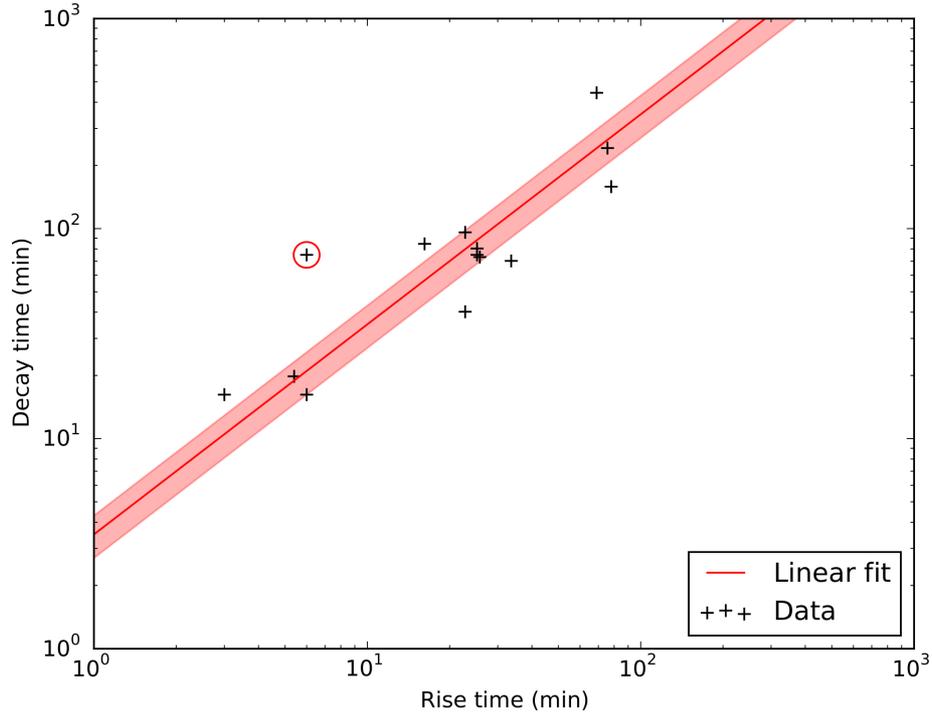}}
\caption{The datapoints show the rise and decay times as calculated for our selection of 14 GLEs. The red line shows a linear fit to the observations, while the red band indicates a $1 \sigma$ (standard deviation) error on the fitted slope. {The data-point corresponding to GLE 48 is circled.}}\label{fig_data_with_fit}
\end{figure*}

Figure \ref{fig_data_with_fit} shows our calculations of the decay and rise times for the 14 selected GLE events as the data points. Although there is still a significant spread in these observed values, a linear trend is discernible. By fitting a linear curve (through the origin) to the data, we find $\tau_\mathrm{d} = (3.5 \pm 0.8) \tau_\mathrm{r}$, where the error is simply taken to be a standard deviation. The regression curve, with its error, is shown on the figure as the red line and shaded region respectively. {All the data-points are fairly well described by the linear fit, with the possible exception of that corresponding to GLE 48 (the circled data-point) which has an exceedingly long decay time. A closer inspection of GLE 48's temporal profile shows a second peak (which might be a second event) some $\approx 4$ hours after the first. It is possible that this second increase influences our calculation of the decay time which was calculated only for the first event.} Simply stated, our analysis {therefore} indicates that the decay time is roughly 3 to 4 times longer than the rise time for a broad range of time-scales containing both impulsive and gradual events. This value differs from that reported by \citet{moraaletal2016}{, {\it i.e.} $\tau_\mathrm{d} \approx 2 \tau_\mathrm{r}$}, although they used a different definition of $\tau_\mathrm{r}$, but similarly also found a linear relationship between $\tau_\mathrm{r}$ and $\tau_\mathrm{d}$. In the next sections, we interpret this seemingly universal dependence in terms of a particle transport model.

\section{Comparison with Theory}

\subsection{The Transport Model}

The propagation of a gyro-tropic distribution of SEPs can be described by the \citet{roelof1969} transport equation (TPE),

\begin{equation}
\label{Eq:TPE}
\frac{\partial f(z,\mu,t)}{\partial t} = - \mu v \frac{\partial f}{\partial z}  - \frac{1-\mu^2}{2L} v \frac{\partial f}{\partial \mu}  + \frac{\partial}{\partial \mu} \left(D_{\mu\mu} (z,\mu) \frac{\partial f}{\partial \mu}
 \right) ,
\end{equation}

\noindent which is valid for motion along a magnetic field directed along $\mathbf{e}_z$. In this expression, $\mu$ is the cosine of the particle's pitch angle, so that the parallel speed becomes $v_{||}=v \mu$, and focussing due to the diverging \citet{parkerhmf} magnetic field is given in terms of the focussing length

\begin{equation}
L^{-1}(z) = \frac{1}{B(z)} \frac{\partial B(z)}{\partial z}
\end{equation}

\noindent and pitch-angle diffusion is included via the pitch-angle diffusion coefficient $D_{\mu \mu}$. Although some analytical solutions of Equation (\ref{Eq:TPE}) do exist (see {\it e.g.}\citet{effielitvi2014}), these are notoriously inaccurate during the rising phase of a SEP event, and as such, we opt to solve  the TPE numerically as outlined in Appendix \ref{Sec:appe_A}.\\

Usually, all transport quantities are specified, not in terms of $z$, but rather in terms of spherical coordinates, and as such, we need to convert between these two coordinate systems, $z \mapsto (r,\phi)$. The unit vector $\mathbf{e}_z$ is defined to point along the magnetic field, $\vec{B}$, so that

\begin{equation}
\mathbf{e}_z  \equiv \frac{\vec{B}}{B} = \cos \Psi \mathbf{e}_r - \sin \Psi \mathbf{e}_{\phi}
\end{equation}

\noindent in terms of spherical coordinates with $\Psi$ being the magnetic field spiral angle, {\it i.e.} the angle between the magnetic field and the radial direction. For a \citet{parkerhmf} magnetic field,

\begin{equation}
\vec{B} = B_0 \left[ \frac{r_0}{r} \right]^2 \left(\mathbf{e}_r - \tan \Psi \mathbf{e}_{\phi}  \right),
\end{equation}

\noindent where $B_0$ is some reference values at $r_0$, we have

\begin{equation}
\tan \Psi = - \frac{B_{\phi}}{B_r} = \frac{\Omega r \sin \theta}{V},
\end{equation}

\noindent which, at Earth in the equatorial plane $(\theta = \pi/2)$, gives $\Psi \approx 45^{\circ}$ when using observed values of the solar wind speed $V\approx 400$ km.s$^{-1}$ and the solar rotation rate $\Omega\approx 25$ days. As the magnitude of $z$ is now defined as the length along the spiral field, we may calculate

\begin{equation}
\label{Eq:def_z}
z(r) = \int_{0}^{r} \mathrm{d}z(r) = \int_{0}^{r} \mathrm{d}r  \sqrt{1 + \tan^2 \Psi} 
\end{equation}

\noindent where we have used 

\begin{equation}
\mathrm{d} \phi = - \frac{\Omega}{V} \mathrm{d}r.
\end{equation}

\noindent For ease, Equation (\ref{Eq:def_z}) is integrated numerically in the present model. Due to the curvature of the magnetic field, we find $z(r=1 \mathrm{ AU}) \approx 1.2 \mathrm{ AU}$ and $z(r=3 \mathrm{ AU}) \approx 5.9 \mathrm{ AU}$.\\

Following \citet{drogeetal2010}, we parametrize the pitch-angle diffusion coefficient as 

\begin{equation}
D_{\mu\mu}(z,\mu) = D(z) \left( 1 - \mu^2 \right) \left\{ \left| \mu \right|^{q-1} + H \right\}
\end{equation}

\noindent where $q=5/3$ is the spectral index of the inertial range of the turbulence power spectrum and $H=0.05$ is a parameter to account for non-linear or dynamical processes which scatter particles though $\mu =0$ \citep[see \emph{e.g.}][]{drogeetal2010}. Here $D(z)$ is a function which we will use to obtain the required magnitude of $D_{\mu \mu}$. In order to compare our results to analytical approximations, and to get a more intuitive feeling for the level of particle scattering, we specify the parallel mean free path $\lambda_{||}$ in the model rather than $D_{\mu\mu}$. These two quantities are related through \citep[][]{Hasselmann}

\begin{equation}
\label{Eq:def_lambda}
 \lambda_{||}(z) = \frac{3v}{8} \int_{-1}^{+1} \frac{\left( 1 - \mu^2 \right)^2}{D_{\mu\mu}(z,\mu)} \mathrm{d}\mu, \nonumber
\end{equation}

\noindent while $\lambda_{||}$ is also related to the effective radial mean free path $\lambda_{rr}$ (in the absence of perpendicular diffusion) through

\begin{equation}
\lambda_{rr}=   \lambda_{||} \cos^2 \Psi.
\end{equation}

\noindent We parametrize $\lambda_{rr}$ as

\begin{equation}
\label{Eq:parametrized}
\lambda_{rr} = \lambda_0 \left[ \frac{r}{r_0} \right]^{\alpha},
\end{equation}

\noindent where $\lambda_0$ is a reference value at $r_0 = 1$ AU. Note that the same parametrization is used in Appendix \ref{Se:appendix_2}, keeping in mind that $\kappa_{rr} = v/3 \lambda_{rr}$, in order to compare with the analytical approximations. By specifying $\lambda_0$ and $\alpha$, we can then follow Equations (\ref{Eq:def_lambda}) - (\ref{Eq:parametrized}) in reverse to estimate $D(z)$ which is the input to the model. The magnitude of $\lambda_0$ can already give us an indication as to the behaviour of SEP intensities: at Earth, $L \approx 1$ AU, and the length of the field line connecting Earth to the Sun is $z \approx 1.2$ AU. Therefore, if $\lambda_0 \ll L, z \approx 1$ AU, we will have very efficient particle scattering and the resulting distribution will be nearly isotropic. If not, {\it i.e.} $\lambda_0 \approx 1$ AU, or larger, the resulting distributions will be highly anisotropic. \\

After solving Equation (\ref{Eq:TPE}) to obtain the distribution function $f(z,\mu,t)$ we can calculate the omni-directional intensity by simply averaging over pitch-angle:

\begin{equation}
F(z,t) =\frac{1}{2} \int_{-1}^{+1}   f(z,\mu,t)  \mathrm{d}\mu,
\end{equation}

\noindent and compare the modelled time profile of $F(z=1.2 \mathrm{ AU},t)$ to that derived from GLE observations. {See also the recent modelling, using a similar approach, by \citet{butikoferetal2016}.}


\subsection{General Modelling Results}

\begin{figure*} 
\centerline{\includegraphics[width=0.99\textwidth,clip=]{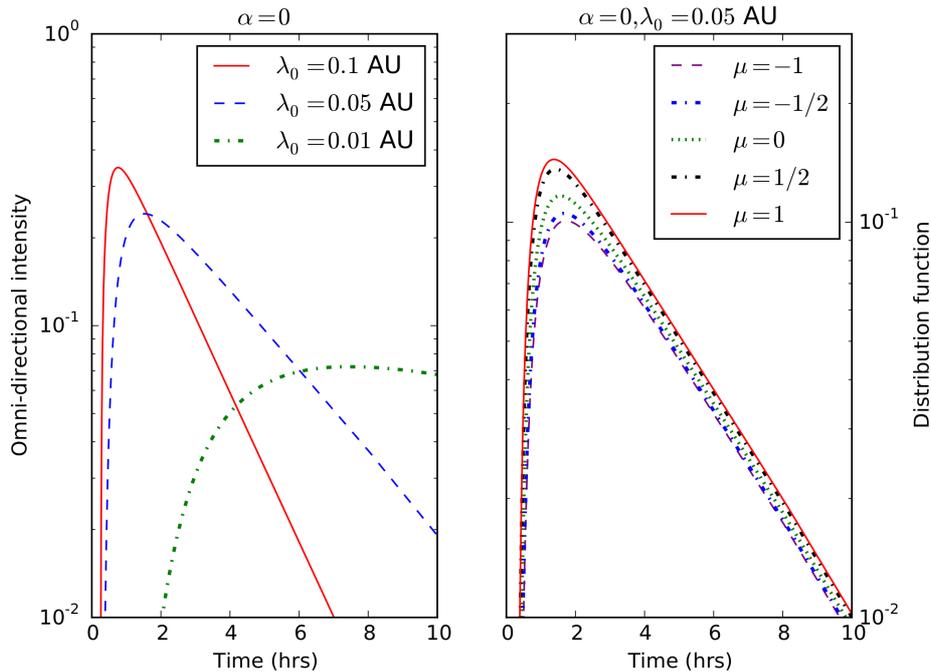}}
\caption{The left panel shows model solutions of the omni-directional intensity for different values of $\lambda_0$ when $\alpha=0$ is kept constant. The right panel shows the distribution function, for $\lambda_0=0.05$ AU and $\alpha = 0$, at different pitch-angles. {All profiles are shown at a radial position of 1 AU.}}\label{fig_model_characteristics}
\end{figure*}

\begin{figure*} 
\centerline{\includegraphics[width=0.5\textwidth,clip=]{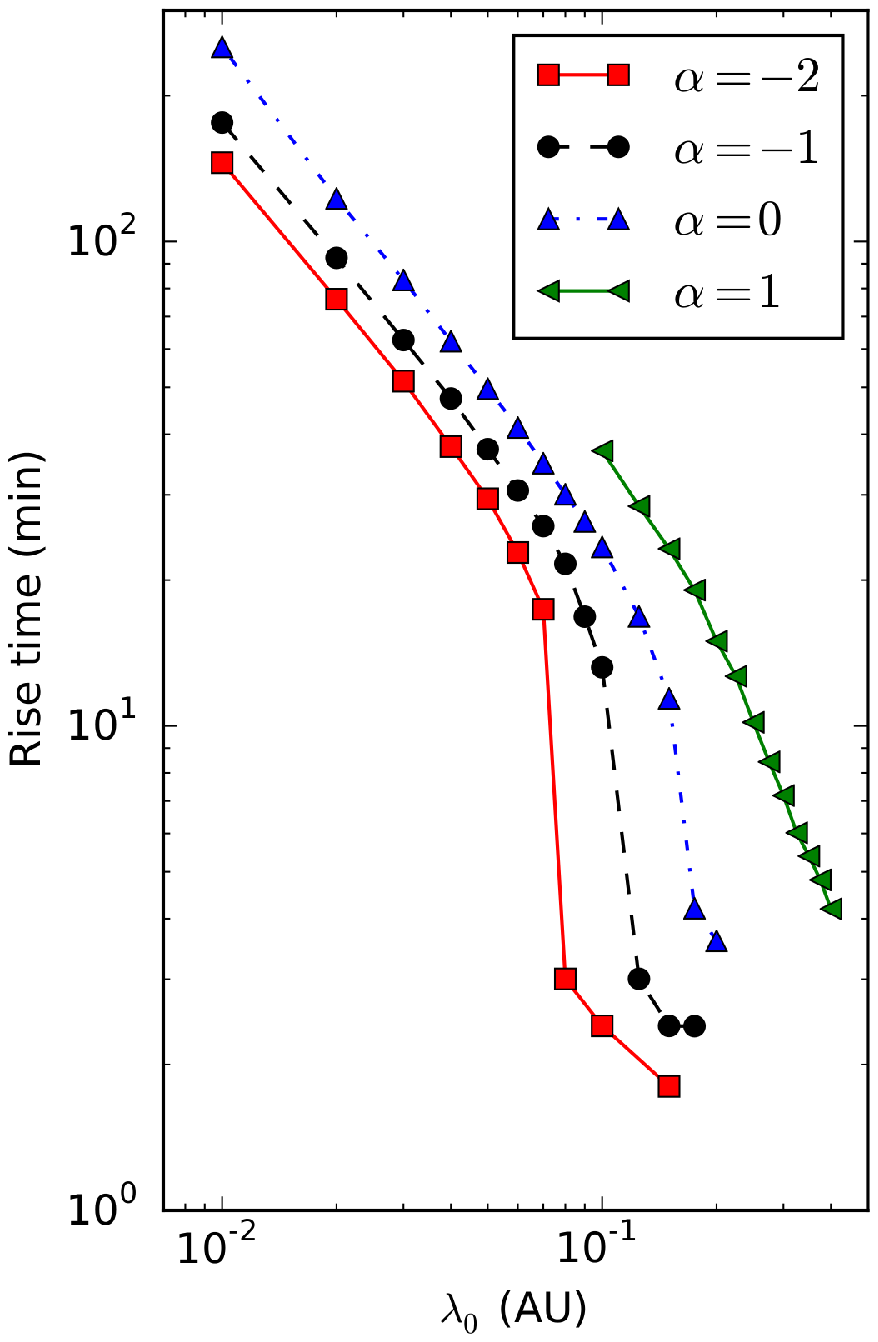}}
\caption{The calculated rise time, as a function of $\lambda_0$, for different values of $\alpha$.}\label{fig_rise_vs_lambda}
\end{figure*}

To be compatible with the GLE observations, we solve Equation (\ref{Eq:TPE}), for 2 GV protons, by specifying isotropic impulsive injection at the inner boundary of the model, $f(z=0.05 \mathrm{ AU}, \mu, t) = \delta (t)$. Examples of the resulting time profiles of $F$, at Earth, are shown in the left panel of Figure \ref{fig_model_characteristics} for a fixed value of $\alpha = 0$ and different values of $\lambda_0$. As expected, the level of particle scattering (quantified through the magnitude of $\lambda_0$) has a large influence on the temporal profiles. For large values of $\lambda_0$ (weak scattering), we note a very fast rise in intensities, followed by a sharp decrease. For smaller values of $\lambda_0$ (stronger scattering), the intensities rise more gradually. We are therefore able to control the temporal profile of the distribution at Earth by adjusting the amount of scattering that the particles experience. Hence, the injection profile (here, simply a delta-function) is not the only controlling factor of the pulse shape of GLEs; interplanetary transport conditions can also have a large influence. In the right panel of Figure \ref{fig_model_characteristics}, we keep $\alpha = 0$ and $\lambda_0 = 0.05$ AU constant and show the distribution function at different pitch-angles. Of course, the particles that stream along the magnetic field, {\it i.e.} these with $\mu = 1$, reach Earth first, and their profile shows a very quick rise ({\it i.e.} a short rise time). Particles mostly gyrating perpendicular to the field ($\mu = 0$), take longer to reach Earth but also have a longer rise time. This is an additional motivation for us to use only the omni-directional intensity to quantify the temporal profile (for both the model and the observations) of the GLE events, thereby removing the dependence of the rise time on pitch-angle, or, in the observational sense, the dependence of the rise time on the geographical position of the NMs.\\

Following our standard definition, we calculate the rise time from the modelled solutions, and present these in Figure \ref{fig_rise_vs_lambda}. As alluded to in Figure \ref{fig_model_characteristics}, larger values of $\lambda_0$ lead to shorter rise times. What is however very interesting, is the linear dependence of $\tau_\mathrm{r}$ for small values of $\lambda_0$; when $\lambda_0$ is less than $\approx 0.1$ AU, $\tau_\mathrm{r}$ decreases linearly with time, with the slope of the linearity dependent on the value of $\alpha$. These small values of $\lambda_0$ must result in fairly isotropic distributions and, as we will show in the next section, this linear dependence is characteristic of the isotropic scenario where pitch-angle scattering is sufficiently strong. As shown in the next section, a similar linear dependence is observed for $\tau_\mathrm{d}$ in this limit.


\subsection{Comparison between Theory and Observations}

\begin{figure*} 
\centerline{\includegraphics[width=0.99\textwidth,clip=]{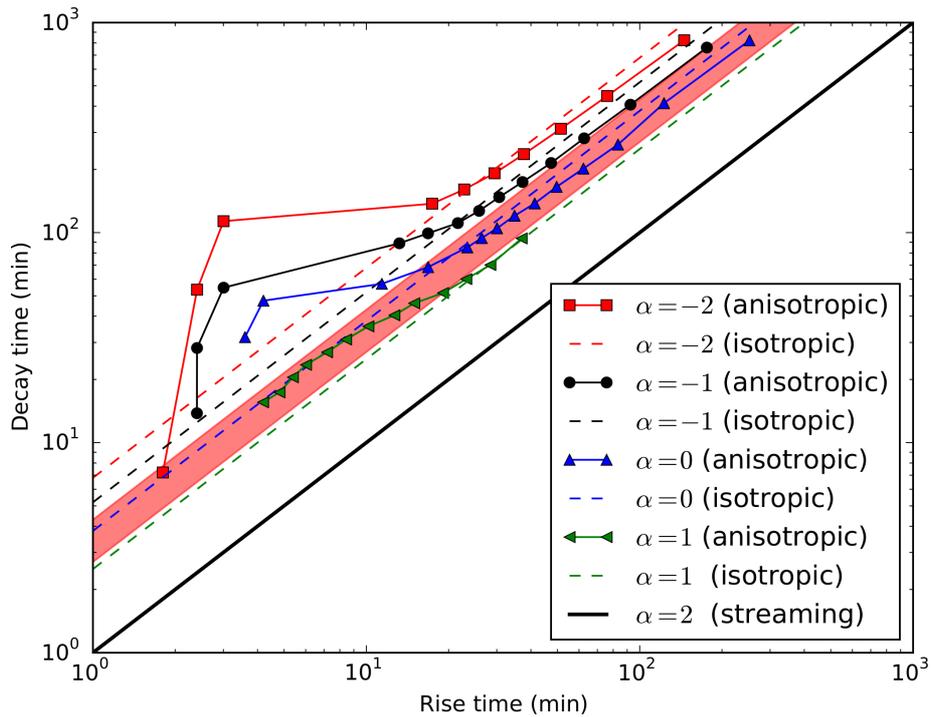}}
\caption{The symbols show the modelled relationship between $\tau_\mathrm{r}$ and $\tau_\mathrm{d}$ for different values of $\alpha$, while the dashed lines are the corresponding analytical approximations discussed in Appendix \ref{Se:appendix_2}. The red band illustrates the relationship as derived from observations in Section \ref{Sec:charactering_the}. }\label{fig_data_with_model}
\end{figure*}

We show the modelled relationship between $\tau_\mathrm{r}$ and $\tau_\mathrm{d}$ in Figure \ref{fig_data_with_model}, as the symbols, for different choices of $\alpha$. As already discussed, by decreasing $\lambda_0$ in the model, both $\tau_\mathrm{r}$ and $\tau_\mathrm{d}$ increase, and in the limit of $\lambda_0 \ll 1$ AU, these increases are linear. The same behaviour is illustrated in Figure \ref{fig_data_with_model} where, for $\tau_\mathrm{r} > 10$ min, a linear relationship is evident between $\tau_\mathrm{r}$ and $\tau_\mathrm{d}$ with the slope dependent on the value of $\alpha$. This linear dependence is indicative of isotropy, and this idea can be tested by comparing the modelled solution to the analytical approximations discussed in Appendix \ref{Se:appendix_2}, which are derived for an isotropic particle distribution, and shown on Figure \ref{fig_data_with_model} as the different dashed lines. Indeed, at larger values of $\tau_\mathrm{r}$ both the model and the approximations follows the same linear trend. Our results thus indicate that for GLEs with $\tau_\mathrm{r} > 10$ min, we may approximate the particle distributions to be isotropic and use the approximations derived in Appendix \ref{Se:appendix_2}. This linear trend of course breaks down below $\tau_\mathrm{r} < 10$ min, as these events are highly anisotropic with $\lambda_0 \approx 1$ AU. No clear dependence is noted for these events.\\

Also shown on Figure \ref{fig_data_with_model} as the red band, is the linear dependence of $\tau_\mathrm{r}$ and $\tau_\mathrm{d}$ as derived form GLE observations in Section \ref{Sec:charactering_the}. We can now explain this linear dependence in terms of particle transport in the isotropic limit. Moreover, a comparison between modelled and observed results indicates the value $\alpha \approx 0$ ($\lambda_{rr}$ independent of radial distance) is consistent with the observations; the possible implications thereof are discussed in the next section.

\section{Summary and Conclusions}

By calculating the rise and decay times for a {small} subset of all GLE events, we find a linear relationship between these two quantities, conveniently summarized as $\tau_\mathrm{d} \approx 3.5 \tau_\mathrm{r}$, {\it i.e.} the decay phase lasts, on average, 3 to 4 times longer than the rising phase of the event. This relationship seems to hold for a large range of $\tau_\mathrm{d}$ and $\tau_\mathrm{r}$ values (almost two orders of magnitude), suggesting that GLEs do not fall into two distinct classes of either impulsive or gradual events, but follow a continuous distribution of impulsive-like or gradual-like events.  It is difficult to imagine how different acceleration mechanisms can lead to such a ``universal" linear relationship, and hence we interpret this result as being indicative of interplanetary transport. We show that interplanetary scattering can significantly affect the temporal profile of GLEs as observed at Earth, in a sense obscuring the initial acceleration profile. In the limit of very effective particle scattering, {\it i.e.} when the resulting distribution is nearly isotropic, both the numerical and analytical solutions reproduce the observed linear trend. We thus conclude that interplanetary transport may have an extremely large effect on the observed pulse shape of GLEs and should not be completely ignored as is mostly done.\\

There are very large inter-event variations in the values of $\tau_\mathrm{d}$ and $\tau_\mathrm{r}$, although the values follow the same general trend. These large variations are most likely related to changes in the level of interplanetary scattering between events (see {\it e.g.} \citet{droge2000}) and can be accounted for in our model by varying, for example, $\lambda_0$. A comparison between the observations and especially the analytical approximations, suggests, however, that a constant radial dependence of $\lambda_{rr} \propto r^0$ (i.e. $\lambda_{rr}$ being independent of radial position) best reproduces the observations for all events. Perhaps rather fortuitously, the same assumption is frequently used in many transport models (see the discussion by \citet{droge2000}), giving us some confidence in our modelling approach and conclusions. \\

{In future, we plan to extend our analyses to include more events, as our 14 selected events only represent $\approx 20\%$ of the total recorded GLEs. However, based on the findings of \citet{moraaletal2015}, who included more events, the qualitative conclusions presented here are not expected to change much. Of particular interest will also be to perform an equivalent study for SEP events that are observed {\it in-situ} by spacecraft and that can be classified more rigorously into impulsive (flare accelerated) and gradual (CME accelerated) events. }

\appendix

\section{Notes on the Numerical Model}
\label{Sec:appe_A}

\begin{figure*} 
\centerline{\includegraphics[width=0.99\textwidth,clip=]{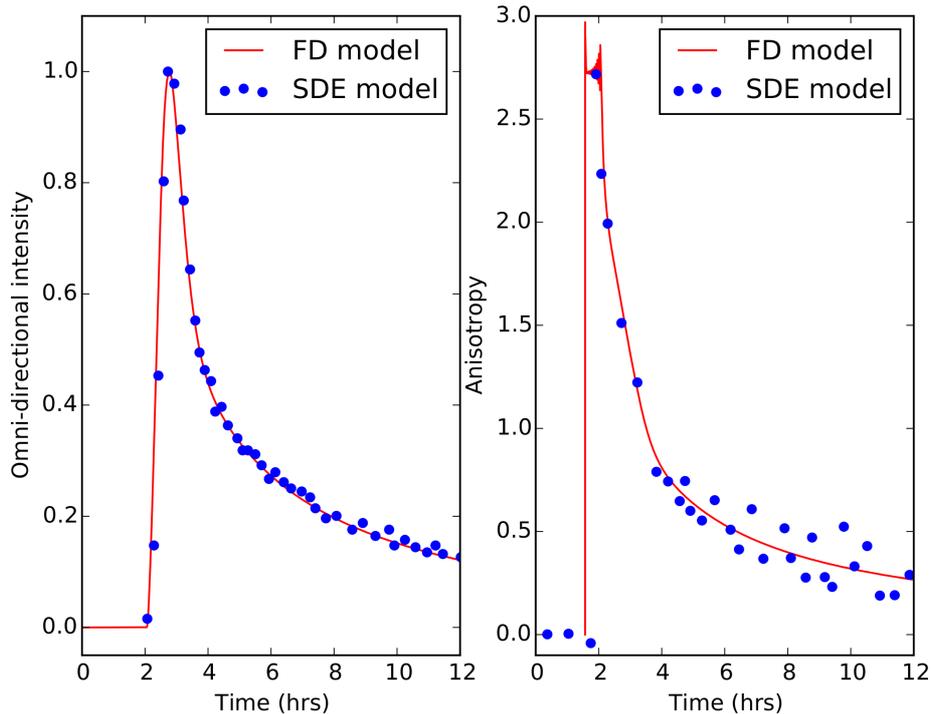}}
\caption{A comparison between our numerical solver (FD model; indicated by the red lines) and that of \citet{drogeetal2010} (SDE model; indicated by the blue symbols). The left panel compares the calculated omni-directional intensity and the right panel the corresponding first-order anisotropy.}\label{figa1}
\end{figure*}

Following \citet{straussfichtner2015}, we integrate Equation (\ref{Eq:TPE}) numerically by first transforming it into a set of three, 1D convection and diffusion equations, using a first-order operator splitting method, to obtain 

\begin{eqnarray}
\frac{1}{3}\frac{\partial f}{\partial t'} + \mu v \frac{\partial  f}{\partial z} &=&  0  \label{Eq:adve_1} \\
\frac{1}{3}\frac{\partial f}{\partial t'} +  \frac{1-\mu^2}{2L} v \frac{\partial f}{\partial \mu}  &=&  0 \label{Eq:adve_2} \\
\frac{1}{3}\frac{\partial f}{\partial t'} &=&  \frac{\partial }{\partial \mu} \left( D_{\mu\mu} \frac{\partial f}{\partial \mu} \right), \label{Eq:diff_equation}
\end{eqnarray}

\noindent where $\mathrm{d}t' = \mathrm{d}t/3$. The diffusion equation (Equation (\ref{Eq:diff_equation})) is solved by an explicit Euler method, while the convection/advection equations (Equations (\ref{Eq:adve_1}) and (\ref{Eq:adve_2})) are solved by a flux limiter corrected upwind scheme. For all equations, flux conserving boundary conditions are applied (see the discussion by \citet{straussfichtner2015}). To validate this modelling approach, we compare our results to those of \citet{drogeetal2010} in this section. These simulations, using exactly the same transport parameters as given by \citet{drogeetal2010}, are for 4 MeV protons with the results shown as a function of time at Earth's position. In Figure \ref{figa1} our model results are labelled as the finite-difference (FD) solutions, while the \citet{drogeetal2010} results, computed by making use of stochastic differential equations (SDEs) are labelled accordingly. The left panel compares the omni-directional intensity and the right the first-order anisotropy, defined as

\begin{equation}
A = 3 \frac{\int_{-1}^{1}  \mu f  \mathrm{d}\mu}{\int_{-1}^{1} f   \mathrm{d}\mu} .
\end{equation}

\noindent The excellent agreement between the two independent models validates our modelling approach while giving us confidence is our calculated temporal profiles.

\section{(Semi-) Analytical Approximations}
\label{Se:appendix_2}

\begin{figure*} 
\centerline{\includegraphics[width=0.5\textwidth,clip=]{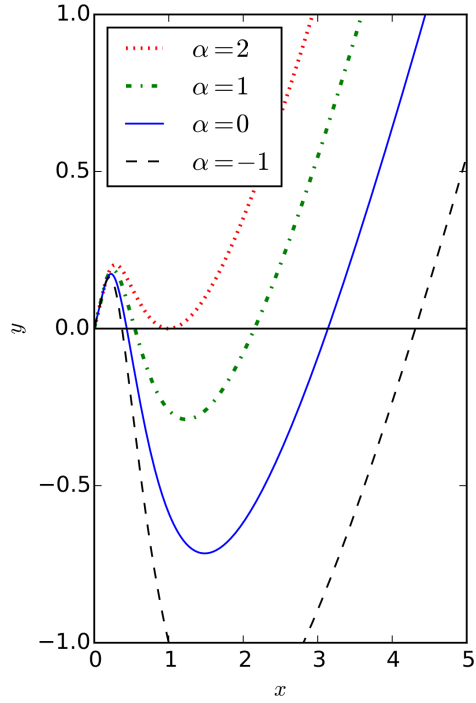}}
\caption{Finding the roots of Equation (\ref{Eq:transcendental_function}) for different values of $\alpha$.}\label{figa2}
\end{figure*}

If the particle distribution can be approximated to be nearly-isotropic, we can follow \citet{moraaletal2016} and obtain $f$ by solving the spherical symmetric diffusion equation

\begin{equation}
\frac{\partial f}{\partial t} = \frac{1}{r^2} \frac{\partial}{\partial r} \left( r^2 \kappa_{rr} \frac{\partial f}{\partial r} \right).
\end{equation}

\noindent Assuming that the effective radial diffusion coefficient, $\kappa_{rr}$, can be parametrized as

\begin{equation}
\kappa_{rr} = \kappa_0 \left[ \frac{r}{r_0} \right]^{\alpha},
\end{equation}

\noindent the solution of $f$ for impulsive injection at $r=0$ and $t=0$ is \citep[][]{dugal1979}

\begin{equation}
f \propto t^{3/(\alpha - 2)} \exp \left[ \frac{-r^2 \left( r_0/r \right)^{\alpha}}{ (2 - \alpha)^2 \kappa_0 t} \right]. \label{Eq:general_solution}
\end{equation}

\noindent The time of peak, or maximum, intensity is evaluated as

\begin{equation}
t_{\max} = \frac{r^2 \left( r_0/r \right)^{\alpha}}{3 (2 - \alpha) \kappa_0},
\end{equation}

\noindent with a corresponding peak intensity of 

\begin{equation}
f_{\max} \propto \left(\exp \left( 1 \right) \cdot t_{\max} \right)^{3/(\alpha - 2)} ,
\end{equation}

\noindent so that Equation (\ref{Eq:general_solution}) may be rewritten, in terms of these quantities, as

\begin{equation}
\frac{f}{f_{\max}} = \left[ \frac{t}{t_{\max}} \exp \left( \frac{t_{\max}}{t} - 1 \right)  \right]^{3/(\alpha - 2)}. \label{Eq:rewritten-solution}
\end{equation}

\noindent This analytical approximation is shown, for a variety of different parameters, and discussed in more detail in \citet{moraaletal2015,moraaletal2016}. Here, however, we are only interested in times where the distribution obtains a half of its maximum value, i.e. we are interested in finding $t^a_{1/2}$ and $t^b_{1/2}$ as defined in Figure \ref{fig_define_times}. By setting $2f=f_{\max}$ in Equation (\ref{Eq:rewritten-solution}), we obtain the required transcendental equation

\begin{equation}
\frac{t_{1/2}}{t_{\max}} = 2 ^{(2 - \alpha)/3} \exp \left( 1 - \frac{t_{\max}}{t_{1/2}}  \right) \label{Eq:transcendental_function}
\end{equation}

\noindent that must be solved numerically to obtain the two values of $t^a_{1/2} < t_{\max}$ and $t^b_{1/2} > t_{\max}$. Note also that this expression is independent of the magnitude of the diffusion coefficient and only on its radial dependence through $\alpha$. Figure \ref{figa2} illustrates how to find the roots (values of $x$ where $y=0$) of Equation (\ref{Eq:transcendental_function}) for different values of $\alpha$ by setting

\begin{eqnarray}
x &:=& \frac{t_{1/2}}{t_{\max}}, \\
y(x, \alpha) & := & x - 2 ^{(2 - \alpha)/3} \exp \left( 1 - \frac{1}{x} \right).
\end{eqnarray}

\noindent For example, $\alpha = 0$ gives $t^a_{1/2} \approx 0.44 t_{\max}$ and $t^b_{1/2} \approx 3.13 t_{\max}$, which, substituted into our definitions of the rise and decay times (see again Section \ref{Sec:charactering_the}) leads to $\tau_\mathrm{d} \approx 3.8 \tau_\mathrm{{r}}$. Our results are summarized as

\begin{eqnarray}
\alpha = -2&:& \tau_\mathrm{{d}} \approx 6.8 \tau_\mathrm{{r}}\\
\alpha = -1&:& \tau_\mathrm{{d}} \approx 5.2 \tau_\mathrm{{r}}\\
\alpha = 0&:& \tau_\mathrm{{d}} \approx 3.8 \tau_\mathrm{{r}}\\
\alpha = 1&:& \tau_\mathrm{{d}} \approx 2.5 \tau_\mathrm{{r}}
\end{eqnarray}

\noindent with the special case of $\tau_\mathrm{{d}} = \tau_\mathrm{{r}}$ for $\alpha = 2$ describing a free-streaming scenario. Note that, for all values of $\alpha$, we obtain a linear relationship between $\tau_\mathrm{{d}}$ and $\tau_\mathrm{{r}}$. These resulting estimates, referred to as the isotropic solutions, are included in Figure \ref{fig_data_with_model} as the dashed lines.

\begin{acks}

This work is based on the research supported in part by the National Research Foundation (NRF) of South Africa (grant no. 106049). Opinions expressed and conclusions arrived at are those of the authors and are not necessarily to be attributed to the NRF. OO acknowledges the support of the post-doctoral programme of the North-West University in South Africa. \emph{This paper is dedicated to the memory of the late Harm Moraal.}

\end{acks}

\begin{acks}[Disclosure of Potential Conflicts of Interest] The authors declare that they have no conflicts of interest. \end{acks}

\end{document}